\documentclass[aps,prl,twocolumn,showpacs,preprintnumbers,
               amsmath,amssymb]{revtex4}
\usepackage{graphicx}
  \usepackage{hyperref}

\newcommand{\be}{\begin{equation}}
\newcommand{\ee}{\end{equation}}
\newcommand{\ba}{\begin{array}}
\newcommand{\bqa}{\begin{eqnarray}}
\newcommand{\eqa}{\end{eqnarray}}
\newcommand{\cO}{{\cal O}}
\newcommand{\Frac}[2]{\frac{\displaystyle #1}{\displaystyle #2}}
\newcommand{\Int}{\displaystyle{\int}}

\begin{document}


\title{
 $O(p^6)$ extension of the
large--$N_C$ partial wave dispersion relations
 }

\author{Z.~H.~Guo, J.J.~Sanz-Cillero~\footnote{Address after October 2007:  Grup de F\'\i sica Te\`orica and IFAE,
Universitat Aut\`onoma de Barcelona, 08193 Barcelona, Spain} and
H.Q.~Zheng }

\affiliation{ Department of Physics, Peking University, Beijing
100871, P.R. China }


\date{\today}

\begin{abstract}
Continuing our previous work~\cite{lnc}, large--$N_C$ techniques and
partial wave dispersion relations are used to discuss $\pi\pi$
scattering amplitudes. We get a set of predictions for $O(p^6)$
low-energy chiral perturbation theory couplings. They  are provided
in terms of the masses and decay widths of scalar and vector mesons.
\end{abstract}
\vskip .5cm


\date{\today}

\maketitle


\section*{Introduction}

Chiral perturbation theory ($\chi$PT) is a powerful tool in the
study of low energy hadron physics. An important issue in $\chi$PT
is the determination of the values of low energy constants (LECs),
which are crucial to make predictions. In addition to an exhaustive
phenomenological discussions about the LECs,
Refs.~\cite{Ecker89}~and~\cite{spin1fields} provided a deeper
theoretical understanding. In these papers, the authors constructed
a phenomenological lagrangian including  the heavy resonances, which
were then integrated out to predict the LECs at tree level in terms
of the resonance couplings.

In a previous paper~\cite{lnc},
we obtained a generalization of the $KSRF$ relation~\cite{KSRF},
a new relation between resonance couplings
and a prediction for the chiral constants $L_2$ and $L_3$~\cite{Polyakov}:
\begin{eqnarray}
&&  \Frac{144 \pi f^2 \overline{\Gamma}_V}{\overline{M}_V^3 }
\, +\, \Frac{32\pi f^2 \overline{\Gamma}_S}{\overline{M}_S^3}\,
=\, 1\, ,
\nonumber \\
&&\Frac{9 \overline{\Gamma}_V}{\overline{M}_V^5}
\left[\alpha_V+6\right]
\, +\, \Frac{2 \overline{\Gamma}_S}{3 \overline{M}_S^5 }
\left[ \alpha_S+6\right] \,
=\, 0\, ,   \nonumber
\nonumber \\
 L_2 \, &=&\, 12 \pi f^4 \Frac{\overline{\Gamma}_V}{\overline{M}_V^5} \, ,
\nonumber \\
 L_3 \, &=&\, 4 \pi f^4
\left( \Frac{2\overline{\Gamma}_S}{3 \overline{M}_S^5} \,
-\,  \Frac{ 9 \overline{\Gamma}_V}{\overline{M}_V^5} \right) \, ,
\end{eqnarray}
where $\overline{\Gamma}_R$ and $\overline{M}_R$ stand, respectively,
for the value of the $R$ resonance width and mass in the chiral limit.
The parameter $\alpha_R$ is given by their $\cO(m_\pi^2)$ correction in the ratio
$\frac{\Gamma_R}{M_R^3}=\frac{\overline{\Gamma}_R}{\overline{M}_R^3}
\left[1+\alpha_R\frac{m_\pi^2}{\overline{M}_R^2}+\cO(m_\pi^4)\right]$.

No particular realization of the resonance lagrangian was considered
in Ref.~\cite{lnc}. While in the lagrangian approach one has to pay
attention to different realizations of the vector
fields~\cite{spin1fields},  all our analyses only rely on general
properties like crossing symmetry and analyticity. Chiral symmetry
was incorporated by matching chiral perturbation theory ($\chi$PT)
at low energies~\cite{chptweinberg,chptoneloop,gasser85}.
In Ref.~\cite{lnc}, we found
that the minimal resonance chiral theory lagrangian~\cite{Ecker89}
was unable to fulfill the high-energy constraints for the partial wave $\pi\pi$-scattering
amplitudes once the matching was taken up to order $p^4$.
Another interesting finding is that
in large $N_C$ limit the [1,1] Pad\'e approximation in SU(3)
$\chi$PT for $\pi\pi$ scatterings means to neglect the left hand
cuts contribution completely~\cite{lhc}, but the understanding to
the latter is very important to accept the $\sigma$ meson even in
the non-linear realization of chiral symmetry~\cite{Xiao00}.
However, in Ref.~\cite{lnc} the $\pi\pi$ scattering was only matched
up to $\cO(p^4)$
This
paper is devoted to extending the discussion up to $O(p^6)$.

\section*{Dispersive analysis}

The $\pi\pi$ scattering amplitude $T(s,t,u)$ admits a decomposition into partial waves of definite
angular momentum $J$~\cite{martin},
\begin{equation}
T(s,t,u)\,\, =\,\, \displaystyle{ \sum_{J} }\, 32\pi (2J+1)\,\, P_J(\cos{\theta})\,\, T_J(s)\, ,
\end{equation}
where every $T_J(s)$ accepts a once-subtracted dispersion relation of the form,
\begin{eqnarray}
&& T_J(s)\, - \, T_J(0) \,= \,
\nonumber \\
&&\,\,  \Frac{s}\pi\Int_{-\infty}^0\Frac{ds' \,
\mbox{Im}T_J(s')}{s'(s'-s)}\,  +\, \Frac{s}{\pi}\Int_{4
m_\pi^2}^\infty  \Frac{ds' \, \mbox{Im}T_J(s')}{s'(s'-s)} \,.
\label{disponce}
\end{eqnarray}
 In general, we will work with amplitudes and partial-waves with
definite isospin, $T(s,t,u)^I$ and $T^I_J(s)$, respectively. We
however quite often in the following omit the indices $I,J$ for
simplicity when no confusion is caused.

At large--$N_C$, the resonances become narrow-width states, allowing
the recovering of the right-hand cut contribution in
Eq.~(\ref{disponce}). In the previous paper \cite{lnc}, we have
demonstrated that the PKU parametrization of S matrix \cite{zheng}
will give the same results in large $N_C$ limit as
Eq.~(\ref{disponce}). The $s$--channel exchange of a resonance $R$
with proper quantum numbers $IJ$ provides for $s>0$ the absorptive
contribution,
\begin{equation} \label{imts}
\mbox{Im}T_J^{I, \rm R}(s)\, =\, \pi \,  \frac{M_{\rm R}\,
\Gamma_{\rm R} }{\rho_{\rm R} }\, \, \delta(s-M_{\rm R}^2)\,,
\end{equation}
where $\rho_{\rm R}=\sqrt{\frac{M_R^2-4m_\pi^2}{M_R^2}}$ and
the subscript $R$ denote the different resonances.

Crossing symmetry relates the right to the left-hand cut through the
expression~\cite{martin},
\begin{eqnarray}
 \label{imtl}
&& \hspace*{-0.5cm}
\mathrm{Im_L}T^{I}_{J}(s) =
\frac{1+(-1)^{I+J}}{s-4m_{\pi}^2}
\sum_{J'}\sum_{I'}(2J'+1)C^{st}_{II'}
\\
&& \hspace*{-0.5cm}\times
\Int_{4m_{\pi}^2}^{4m_{\pi}^2-s}dt
P_J(1+\frac{2t}{s-4m_{\pi}^2})  P_{J'}(1+\frac{2s}{t-4m_{\pi}^2})
\mathrm{Im_R}T^{I'}_{J'}(t)  , \nonumber
\end{eqnarray}
with $P_n(x)$ the Legendre polynomials. The crossing matrix is also given by~\cite{martin}
\begin{equation}
C^{(st)}_{II'}\, =\, \left(\begin{array}{rrr} 1/3 & 1 & 5/3
 \\ 1/3 & 1/2 & -5/6
 \\ 1/3 & -1/2 & 1/6
\end{array}\right)\, .
\end{equation}
Hence,  the imaginary part of $T^I_J(s)$ for $s<0$ produced by the
crossed-channel resonance ($R$) exchange  is given by
\begin{eqnarray} \label{imtt}
&&\mathrm{Im}T^{I, \rm L}_{J}(s)  =
\nonumber \\
&& \,\, \theta(-s-M_{\rm R}^2+4m_\pi^2)
\times  \frac{1+(-1)^{I+J}}{s-4m_{\pi}^2} (2J'+1)C^{st}_{II'}
\\
&& \,\, \times  P_J(1+\frac{2 M_{\rm R}^2}{s-4m_{\pi}^2}) \,
P_{J'}(1+\frac{2s}{M_{\rm R}^2-4m_{\pi}^2}) \,\, \frac{\pi\, M_{\rm
R}\, \Gamma_{\rm R}}{\rho_{\rm R} }\, .\nonumber
\end{eqnarray}

Putting the different imaginary parts together, it is then possible to
calculate the right and left-hand cut integrals:
\begin{eqnarray}
T^{\rm sR}(s)  &=& \frac{s}{\pi}\int_{4 m_\pi^2}^\infty  \frac{ds'
\, \mbox{Im}T^{\rm R}(s') }{s'(s'-s)} \,,
\\
T^{\rm tR}(s)  &=& \frac{s}\pi\int_{-\infty}^0\frac{ds'
\mbox{Im}T^{\rm R}(s') }{s'(s'-s)} \,,
\end{eqnarray}
where these expressions only depend on the mass and width of the
resonances. The precise results for $T^{\rm sR}$ and $t^{\rm tR}$,
with $R=S,V$,  are given in Ref.~\cite{lnc}.

We consider now the low energy limit where the $\pi\pi$ scattering
is described by $\chi$PT which determines  the left-hand side of
Eq.~(\ref{disponce}). For convenience, the dispersion relation is
rewritten  in the way, \be \label{disponce1} T^{\chi PT}(s)-T^{\chi
PT}(0) = T^{tR}(s) + T^{sR}(s)\,, \ee where the  $l.h.s.$  only
contains $\chi PT$ couplings and the  $r.h.s.$  only contains
resonances parameters. Comparing the different terms of the chiral
expansion on both sides, one gets the low-energy constants (LECs) in
terms of parameters of resonances and some other useful relations.

The $\pi\pi$ scattering amplitude is determined by the function
$A(s,t,u)$,
\begin{equation}
\begin{array}{l}
A\left[\pi^a(p_1)+\pi^b(p_2)\to \pi^c (p_3)+\pi^d(p_4)\right]
\,\, =\,\,
\\
\\
\quad  \delta^{ab}\delta^{cd} A(s,t,u)\,
+\, \delta^{ac}\delta^{bd} A(t,u,s)\,
+\, \delta^{ad}\delta^{bc} A(u,t,s)\, ,
\end{array}
\end{equation}
which is given up to $O(p^6)$ in Ref.~~\cite{op61}.
Since we are interested in the $m_\pi$ dependence of the amplitude, we express
the amplitude explicitly in terms of LECs, momenta  and masses:
\begin{eqnarray}
\label{op6amp}
&&
A(s,t,u)\, =\,  \frac{s-m_\pi^2}{f^2}
+\frac{16m_\pi^4}{f^4}\left(L_2+L_3+L_8-\frac{1}{2} L_5 \right) \nonumber
\\ &&
\,
- \frac{16 m_\pi^2s}{f^4}(L_2+L_3)
+\frac{2s^2}{f^4}(2L_3+3L_2)
+\frac{2(t-u)^2}{f^4}L_2
\nonumber
\\ &&
\quad
\frac{16 m_\pi^6}{f^6}(- 8 L_5^2+ 32 L_8 L_5 - 32 L_8^2)
+\frac{m_\pi^6}{f^6} \left(r_1+ 2 r_f\right)  \nonumber
\\ &&
\quad
+\frac{m_\pi^4s}{f^6} \left(r_2- 2 r_f\right)
\, + \,  \frac{m_\pi^2s^2}{f^6}r_3 \nonumber
\\ &&
\quad
+ \frac{m_\pi^2(t-u)^2}{f^6}r_4
+\frac{s^3}{f^6}r_5
+\frac{s(t-u)^2}{f^6}r_6 \nonumber
\\ &&
\end{eqnarray}
with $s=(p_1+p_2)^2,\, t=(p_1-p_3)^2,\, u=(p_1-p_4)^2=4m_\pi^2-s-t$,
and where we have used the chiral expansion of  the pion decay constant $f_\pi$
up to $\cO(p^6)$~\cite{op61,f2loop}:
\begin{equation} \label{fpi}
\hspace*{-0.25cm}
f_\pi = f\left[1+\frac{4L_5m_\pi^2}{f^2}
+(32L_5^2-64L_8L_5+r_f)\frac{m_\pi^4}{f^4}+\cO(m_\pi^6)\right].
\end{equation}
In both expressions, only the leading terms in the $1/N_C$ expansion
are kept. Following the notation in the former work~\cite{lnc},
the large--$N_C$ $\cO(p^4)$  $SU(2)$ LECs have been expressed in
terms of  $SU(3)$ constants~\cite{gasser85}.

The isospin amplitudes are given by the combinations
\bqa
T(s,t,u)^{I=0}&=&3 A(s,t,u)+A(t,s,u)+A(u,s,t) \,, \nonumber \\
T(s,t,u)^{I=1}&=&A(t,s,u)-A(u,s,t) \,, \nonumber \\
T(s,t,u)^{I=2}&=&A(t,s,u)+A(u,s,t) \,. \eqa Finally, in order to get
amplitudes with definite angular momentum, one performs  the
 partial wave projection,
 \be T(s)^I_J =
\frac{1}{32\pi}\frac{1}{s-4m_\pi^2}
 \Int^{0}_{4m_\pi^2-s}
 \hspace*{-0.25cm}
 P_J(1+\frac{2t}{s-4m_\pi^2}) T(s,t,u)^I dt\  .
\ee This yields the $\chi$PT results for different partial-wave
amplitudes up to $\cO(p^6)$:
\begin{enumerate}
\item{$IJ=00$ channel}
\bqa \label{chpt00} &&l.h.s. = \frac{s}{16\pi f^2}-
\frac{10L_2+5L_3}{3\pi f^4}m_\pi^2s \nonumber \\&&
-\frac{-3r_2+8r_3+32r_4+36r_5+4r_6+6r_f}{48\pi f^6}m_\pi^4s
\nonumber \\&& +\frac{25L_2+11L_3}{24\pi
f^4}s^2+\frac{11r_3+17r_4+18r_5+10r_6}{96\pi f^6}m_\pi^2s^2
\nonumber \\ && +\frac{15r_5-5r_6}{192\pi f^6}s^3  ,
\eqa

\item{$IJ=11$ channel}
\bqa \label{chpt11} &&l.h.s. =\frac{s}{96\pi f^2}+ \frac{L_3}{6\pi
f^4}m_\pi^2s \nonumber \\&&
+\frac{5r_2+40r_3-80r_4+216r_5-24r_6-10r_f}{480\pi f^6}m_\pi^4s
\nonumber
\\&&+\frac{-L_3}{24\pi f^4}s^2-\frac{5r_3-15r_4+54r_5+14r_6}{480\pi
f^6}m_\pi^2s^2
\nonumber \\ &&
+\frac{3r_5+3r_6}{320\pi f^6}s^3 ,
\eqa

\item{$IJ=20$ channel}
\begin{eqnarray}
\label{chpt20} &&l.h.s. = -\frac{s}{32\pi f^2}- \frac{8L_2+L_3}{6\pi
f^4}m_\pi^2s \nonumber \\&&
-\frac{3r_2+16r_3+40r_4+72r_5+56r_6-6r_f}{96\pi f^6}m_\pi^4s
\nonumber
\\&&+\frac{5L_2+L_3}{12\pi f^4}s^2+\frac{r_3+7r_4+9r_5+17r_6}{48\pi
f^6}m_\pi^2s^2
\nonumber \\ &&
-\frac{3r_5+11r_6}{192\pi f^6}s^3 .
\end{eqnarray}
\end{enumerate}
where $l.h.s.$ means the left hand side of Eq.~(\ref{disponce1}).

For the $r.h.s. $ of Eq.~(\ref{disponce1}), a similar chiral
expansion is performed up to $\cO(p^6)$:
\begin{enumerate}

\item{$IJ=00$ channel}
\bqa \label{res00s}
&&T^{sR}=\frac{\Gamma_S}{M_S^3}s+
\frac{2\Gamma_S}{M_S^5}m_\pi^2s+\frac{6\Gamma_S}{M_S^7}m_\pi^4s
+\frac{\Gamma_S}{M_S^5}s^2
\nonumber
\\ && \qquad\qquad
+\frac{2\Gamma_S}{M_S^7}m_\pi^2s^2+\frac{\Gamma_S}{M_S^7}s^3+\cO(p^8),\nonumber
\\&&
\eqa
\bqa
\label{res00t}
&&T^{tR}=\frac{-\Gamma_S}{3M_S^3}s-\frac{22\Gamma_S}{9M_S^5}m_\pi^2s-
\frac{122\Gamma_S}{9M_S^7}m_\pi^4s+\frac{9\Gamma_V}{M_V^3}s
\nonumber \\
&&
+\frac{74\Gamma_V}{M_V^5}m_\pi^2s
+\frac{446\Gamma_V}{M_V^7}m_\pi^4s
+ \frac{2\Gamma_S}{9M_S^5}s^2
+\frac{22\Gamma_S}{9M_S^7}m_\pi^2s^2
\nonumber
\\&&
-\frac{\Gamma_S}{6M_S^7}s^3-\frac{4\Gamma_V}{M_V^5}s^2-\frac{46\Gamma_V}{M_V^7}m_\pi^2s^2
+\frac{5\Gamma_V}{2M_V^7}s^3 +\cO(p^8);\nonumber \\&& \eqa

\item{$IJ=11$ channel}
\bqa \label{res11s}
&&T^{sR}=\frac{\Gamma_V}{M_V^3}s+
\frac{2\Gamma_V}{M_V^5}m_\pi^2s+\frac{6\Gamma_V}{M_V^7}m_\pi^4s
+\frac{\Gamma_V}{M_V^5}s^2
\nonumber \\ &&
+\frac{2\Gamma_V}{M_V^7}m_\pi^2s^2+\frac{\Gamma_V}{M_V^7}s^3+\cO(p^8)
,\nonumber \\&& \eqa
\bqa \label{res11t}
&&T^{tR}=\frac{\Gamma_S}{9M_S^3}s+\frac{10\Gamma_S}{9M_S^5}m_\pi^2s+
\frac{326\Gamma_S}{45M_S^7}m_\pi^4s+\frac{\Gamma_V}{2M_V^3}s
\nonumber
\\&&
+\frac{\Gamma_V}{M_V^5}m_\pi^2s
-\frac{37\Gamma_V}{5M_V^7}m_\pi^4s
-\frac{\Gamma_S}{9M_S^5}s^2-\frac{64\Gamma_S}{45M_S^7}m_\pi^2s^2
\nonumber
\\&&
+\frac{\Gamma_S}{10M_S^7}s^3
+\frac{\Gamma_V}{2M_V^5}s^2+\frac{38\Gamma_V}{5M_V^7}m_\pi^2s^2-\frac{11\Gamma_V}{20M_V^7}s^3
+\cO(p^8) ; \nonumber \\&& \eqa

\item{$IJ=20$ channel}
\bqa
\label{res20s}
&& T^{sR}=0 ,
\eqa
\bqa
\label{res20t}
&&T^{tR}=-\frac{\Gamma_S}{3M_S^3}s-\frac{22\Gamma_S}{9M_S^5}m_\pi^2s-
\frac{122\Gamma_S}{9M_S^7}m_\pi^4s-\frac{9\Gamma_V}{2M_V^3}s
\nonumber
\\&&
-\frac{37\Gamma_V}{M_V^5}m_\pi^2s
-\frac{223\Gamma_V}{M_V^7}m_\pi^4s
+\frac{2\Gamma_S}{9M_S^5}s^2+\frac{22\Gamma_S}{9M_S^7}m_\pi^2s^2
\nonumber
\\&&
-\frac{\Gamma_S}{6M_S^7}s^3
+\frac{2\Gamma_V}{M_V^5}s^2+\frac{23\Gamma_V}{M_V^7}m_\pi^2s^2-\frac{5\Gamma_V}{4M_V^7}s^3+\cO(p^8)
\, ;\nonumber
\\&&
\eqa
\end{enumerate}
where only the lightest  multiplet of vector and scalar resonances
is taken into account,  respectively denoted by the subscripts $V$
and $S$.

The masses $M_R$ and decay widths $\Gamma_R$ in
Eqs.~(\ref{res00s})--(\ref{res20t}) denote the physical ones at
large--$N_C$. They carry an implicit $m_\pi^2$ dependence that we
parameterize in the form
\begin{equation}
\label{beta}
\frac{\Gamma_R}{M_R^5}=\frac{\overline{\Gamma}_R}{\overline{M}_R^{5}} \left[1+
\beta_R\frac{m_\pi^2}{\overline{M}_R^{2}}+\cO(m_\pi^4)\right], \ee \be
\label{alphagamma}
\frac{\Gamma_R}{M_R^3}=\frac{\overline{\Gamma}_R}{\overline{M}_R^{3}} \left[1+
\alpha_R\frac{m_\pi^2}{\overline{M}_R^{2}}+\gamma_R\frac{m_\pi^4}{\overline{M}_R^{4}}+\cO(m_\pi^6)\right],
\end{equation}
where $\overline{M}_R$ and $\overline{\Gamma}_R $ are the chiral
limit of $M_R$ and $\Gamma_R$, respectively. Notice that
$\overline{\Gamma}_R$ and $\overline{M}_R$ were denoted as
$M_R^{(0)}$ and $\Gamma_R^{(0)}$ in Ref.~\cite{lnc}.

After expanding the resonance contributions on the $r.h.s.$ of
Eq.~(\ref{disponce1}) in powers of $s$ and $m_\pi^2$, it is possible
to perform a matching with $\chi$PT. Ref.~\cite{lnc} was devoted to
the analysis of the constraints derived from $\chi$PT at $\cO(p^2)$
and $\cO(p^4)$.   The present work studies the relations that stem
from the matching at $\cO(p^6)$
\begin{enumerate}
\item{$IJ=00$ channel}

\bqa  \label{rel000} &&
\frac{3 r_2 - 8 r_3 -32 r_4 -36 r_5 -4 r_6 -6 r_f}{48\pi f^6}=
\nonumber\\&&
\qquad
\frac{\overline{\Gamma}_{\rm S}}{\overline{M}_{\rm S}^7}
\left( -\frac{68}{9} - \frac{4\beta_{\rm S}}{9} +\frac{2 \gamma_{\rm S}}{3} \right)
\nonumber \\&&
\qquad
+\frac{\overline{\Gamma}_{\rm V}
}{\overline{M}_{\rm V}^7}  \left( 446 + 74 \beta_{\rm V} + 9\gamma_{\rm V}\right)
\, ,
\eqa

\bqa \label{rel001} &&
\frac{ 11 r_3 + 17 r_4 + 18 r_5 + 10 r_6   }{96\pi f^6} =
 \nonumber\\&&
\quad \frac{\overline{\Gamma}_{\rm S}}{\overline{M}_{\rm S}^7}
\left(  \frac{40}{9}+\frac{11\beta_{\rm S}}{9}   \right)
+\frac{\overline{\Gamma}_{\rm V}}{\overline{M}_{\rm V}^7}
\left(    -46 -4 \beta_{\rm V} \right)
\, , \nonumber \\
\eqa

\bqa \label{rel002} \frac{15 r_5-5 r_6}{192\pi f^6}\, =\,
\frac{5 \overline{ \Gamma}_{\rm S}}{6 \overline{M}_{\rm S}^7}
\, +\, \frac{5\overline{\Gamma}_{\rm V}}{2 \overline{M}_{\rm V}^7} \, .
\eqa

\item{$IJ=11$ channel}
\bqa \label{rel110}
&& \frac{5 r_2 + 40 r_3- 80 r_4 +216 r_5 -24 r_6-10 r_f}{480 \pi f^6}=
\nonumber
\\ &&
\qquad
\frac{\overline{\Gamma}_{\rm S}}{\overline{M}_{\rm S}^7}
\left(   \frac{326}{45} + \frac{10 \beta_{\rm S}}{9}+\frac{\gamma_{\rm S}}{9}  \right)
\nonumber \\ &&
\qquad
+\frac{\overline{\Gamma}_{\rm V}}{\overline{M}_{\rm V}^7}
\left( -\frac{7}{5}+3 \beta_{\rm V}  +\frac{3 \gamma_{\rm V}}{2} \right)\, ,
\eqa

\bqa  \label{rel111} &&  \frac{-5 r_3 + 15 r_4 -54
r_5 -14 r_6}{480\pi f^6} =
\nonumber \\ &&
\quad
\frac{\overline{\Gamma}_{\rm
S}}{\overline{M}_{\rm S}^7}
\left(-\frac{\beta_{\rm
S}}{9}-\frac{64 }{45}  \right) +\frac{\overline{\Gamma}_{\rm
V}}{\overline{M}_{\rm V}^7}
\left(  \frac{48}{5} +\frac{3\beta_{\rm V}}{2}        \right) \, ,  \nonumber \\
\eqa

\bqa\label{rel112} && \frac{ 3 r_5 + 3 r_6  }{320\pi f^6}\, =\,
\frac{\overline{\Gamma}_{\rm S}}{10 M_{\rm S}^7}\,
+\, \frac{9 \overline{\Gamma}_{\rm V}}{20 \overline{M}_{\rm V}^7}   \, . \eqa

\item{$IJ=20$ channel}

 \bqa \label{rel200}
 &&  \frac{ - 3 r_2 -16 r_3 -40 r_4 -72 r_5 - 56 r_6 + 6 r_f     }{96\pi f^2}=
 \nonumber\\&&
\qquad
  -\, \frac{\overline{\Gamma}_{\rm S}}{\overline{M}_{\rm S}^7}
\left(  \frac{122}{9} +\frac{22\beta_{\rm S}}{9} +\frac{\gamma_{\rm S}}{3}      \right)
\nonumber \\ &&
\qquad
   \, -\, \frac{\overline{\Gamma}_{\rm V}}{\overline{M}_{\rm V}^7}
\left(  223 + 37 \beta_{\rm V} +\frac{9 \gamma_{\rm V}}{2}      \right)   \, ,
\eqa

\bqa \label{rel201} &&
    \frac{r_3+ 8 r_4+9 r_5 +17 r_6}{96 \pi f^6}=
\nonumber\\&&
\quad
 \frac{\overline{\Gamma}_{\rm S}}{\overline{M}_{\rm S}^7}
 \left( \frac{22}{9} +\frac{2\beta_{\rm S}}{9}       \right)
 \, +\,  \frac{\overline{\Gamma}_{\rm V}}{\overline{M}_{\rm V}^7}
\left(   23 + 2\beta_{\rm V}    \right) \,  ,
\eqa

\bqa \label{rel202}  \frac{- 3 r_5 -11 r_6}{192\pi f^6}=\,
\, - \, \frac{\overline{\Gamma}_{\rm S}}{6 \overline{M}_{\rm S}^7}
\, - \,  \frac{5 \overline{\Gamma}_{\rm V}}{4 \overline{M}_{\rm V}^7} \, .
\eqa

\end{enumerate}
Eqs.(\ref{rel000}), (\ref{rel110}) and (\ref{rel200}) refer to the
matching of the terms $\cO(m_\pi^4 s)$. Eqs.~(\ref{rel001}),
(\ref{rel111}) and (\ref{rel201}) correspond to the $\cO(m_\pi^2
s^2)$ terms. Eqs.~(\ref{rel000}),~(\ref{rel110})~and~(\ref{rel200})
provide the matching at $\cO(s^3)$.

It is remarkable that the system of nine equations for six unknowns
($r_{i}$, with $i=f,2...6$) is actually compatible. The $\cO(s^3)$
relations determine $r_5$ and $r_6$. After that, it is then possible
to extract $r_3$ and $r_4$  from the $\cO(m_\pi^2 s^2)$ equations.
Finally, using these values, one can extract the combination $r_2-2
r_f$ from the $\cO(m_\pi^4 s)$ constraints. The LECs always appear
in this particular combination, avoiding an independent
determination of $r_2$ and $r_f$. This yields the predictions: \bqa
\label{r2rf} & &  \hspace*{-1cm} r_2-2r_f= \nonumber \\&& \frac{64
\pi f^6 \overline{\Gamma}_S}{\overline{M}_S^7} \left( 1+
\frac{\beta_{\rm S}}{3} +\frac{\gamma_{\rm S}}{6}   \right)
\nonumber \\ && +\frac{\pi f^6
\overline{\Gamma}_V}{\overline{M}_V^7} \left( 7584 + 1248 \beta_{\rm
V} + 144 \gamma_{\rm V}       \right) \eqa \be \label{r3}
r_3=\frac{64 \pi f^6 \overline{\Gamma}_S}{3\overline{M}_S^7} \left(
1+\frac{\beta_{\rm S}}{2}\right) \, - \, \frac{768 \pi f^6
\overline{\Gamma}_V}{\overline{M}_V^7} ( 1 + \frac{3  \beta_{\rm
V}}{32} )  \, , \ee \be \label{r4}
 r_4=\frac{192 \pi f^6 \overline{\Gamma}_V}{\overline{M}_V^7}
 \left( 1 +\frac{\beta_{\rm V}}{8}\right)
\, ,
\ee
\be
\label{r5}
r_5=\frac{32 \pi f^6 \overline{\Gamma}_S}{3\overline{M}_S^7}+\frac{36 \pi f^6 \overline{\Gamma}_V}{\overline{M}_V^7}
\, ,
\ee
\be \label{r6}
r_6=\frac{12 \pi f^6 \overline{\Gamma}_V}{\overline{M}_V^7} \,  .
\ee

\section*{An example of $\cO(p^6)$ coupling determination}

The authors of Ref.~\cite{op61} provide an estimate of the $\cO(p^6)$
LECs $r_i$  in terms of resonances couplings,
where they consider a phenomenological lagrangian
including one multiplet of vector and scalar resonances.
The vector interaction is given by
\be
\label{lv}
\mathcal{L}_V=
-i\frac{g_V}{2\sqrt{2}}\langle \hat V_{\mu\nu}[u^\mu,u^\nu] \rangle
+f_\chi \langle \hat V_{\mu}[u^\mu,\chi_{-}] \rangle\, ,
\ee
and for the scalar,
\be
\label{ls}
\mathcal{L}_S= c_d \langle S u^\mu u_\mu\rangle \, +\, c_m\langle S\chi_+\rangle\,+
\widetilde{c_d}S_1 \langle u^\mu u_\mu\rangle \, +\, \widetilde{c_m} S_1\langle \chi_+\rangle\,.
\ee
where $\langle ...\rangle$ is short for trace in flavour space
and the tensors $u^\mu,\, \chi_\pm$ introduce the chiral Goldstones.
For further details on the notations, see Ref.~\cite{op61} and references therein.
At large--$N_C$, the $SU(3)$ singlet and octet states become degenerate and one has
$\widetilde{c}_d=c_d/\sqrt{3},\, \widetilde{c}_m=c_m/\sqrt{3},
\, M_{S_1}=M_S$~\cite{Ecker89}.
Using this lagrangian, the authors computed the
contributions to the  $\pi\pi$ scattering from resonance exchanges and provided
a set of values for the LECs $r_i$~\cite{op61}.

As an example of our method, we will rederive  their result.
In order to do that, in a first step, we will neglect the
wave-function renormalizations $Z_R$ and $Z_\pi$,
and only the resonance exchange contribution will be considered,
as it was done in Ref.~\cite{op61}.
At large--$N_C$, the meson wave functions get renormalized
if there are tree-level tadpole diagrams that connect the scalar meson
field to the vacuum~\cite{juanjo,Kpi-SFF}.
After recovering the results in Ref.~\cite{op61},
we will compute the LECs including also the effect of $Z_\pi$ and $Z_R$
and their impact on the numerical estimates will be analyzed.

We need first to calculate the $R\to \pi\pi$ decay
widths corresponding to this lagrangian.
Ignoring the wave-function renormalizations, one gets
\begin{eqnarray}
\label{gammav}
\Gamma_V&=&\frac{g_V^2 M_V^5 \, \rho_{\rm V}^3 }{48\pi f^4}
\left[1 +\Frac{4\sqrt2 f_\chi}{g_V}\, \Frac{m_\pi^2}{M_V^2}\right]^2\,,
\\
\label{gammas}
\Gamma_S &=&
\frac{3 c_d^2 M_S^3\, \rho_{\rm S} }{16\pi f^4}
\left[ 1 + \Frac{2(c_m-c_d)}{c_d}\, \Frac{m_\pi^2}{M_S^2}\right]^2\,,
\end{eqnarray}
where the subscript $S$ denote the $SU(2)$ singlet
$\sigma=\sqrt{\frac{2}{3}}S_0-\sqrt{\frac{1}{3}} S_8  \sim\frac{1}{\sqrt{2}}(\bar{u}u +\bar{d}d)$.
The large--$N_C$ resonances masses are $m_\pi$-independent within this model, i.e.,
$M_{\rm R}=\overline{M}_{\rm R}$.

With the above expressions of $\Gamma_V$ and $\Gamma_S$, we can get
the parameters $\alpha_R$, $\beta_R$ and $\gamma_R$  defined in
Eq.~(\ref{beta})~and~(\ref{alphagamma})
\begin{eqnarray}
\label{av}
\alpha_V\, =\, \beta_V\, &=& \, \Frac{8\sqrt2 f_\chi}{g_V}-6\,,
\\
\gamma_V \, &=& \, \frac{32f_\chi^2}{g_V^2} - \frac{ 48 \sqrt{2} f_\chi}{g_V} +6 \,,
\\
\alpha_S\, =\, \beta_S \, &=& \, \Frac{4c_m}{c_d}-6\,,
\\
\label{gs}
\gamma_S\, &=& \, 10-\frac{16c_m}{c_d}+\frac{4c_m^2}{c_d^2}\,.
\end{eqnarray}

Using Eqs.~(\ref{r2rf})--(\ref{r6}), one gets the predictions on
$\cO(p^6)$ LECs   in terms of the resonance large--$N_C$ parameters
$g_V$, $f_\chi$, $c_d$ and $c_m$:
\begin{equation}
\label{r2}
r_2-2 r_f \, =\, 20 a_V + 16 b_V+3 c_V
+\Frac{ 8 f^2\, \left( c_m-c_d\right)^2}{M_S^4}  \, ,
\end{equation}
\begin{equation}
r_3=-7 a_V - 3 b_V +\Frac{8 f^2c_d(c_m-c_d)}{M_S^4}\,,
\end{equation}
\begin{equation}
r_4=a_V+b_V\,,
\end{equation}
\begin{equation}
r_5= \Frac{3}{4} a_V +\frac{2f^2c_d^2}{M_S^4}\,,
\end{equation}
\begin{equation}
r_6=\Frac{1}{4} a_V\,,
\end{equation}
with $a_V\equiv g_V^2 f^2/M_V^2$, $b_V\equiv 4 \sqrt{2} f_\chi g_V
f^2/M_V^2$, $c_V\equiv 32 f_\chi^2 f^2 /M_V^2$.
If one neglects the wave-function renormalization
and the tadpole effects then the pion decay constant is given by $f_\pi=f$
and therefore $r_f=0$.
Taking this into account, we get a set of predictions for LECs $r_2, \, ... r_5$,
in complete agreement with the results in Ref.~\cite{op61}.

However, all the former results ignored
the effects of the scalar tadpole~\cite{juanjo,Kpi-SFF}.
The term $c_m\langle S \chi_+\rangle$
connects the scalar field to the vacuum, inducing
a pion wave-function renormalization  and
a more complicate relation between $m_\pi$ and the quark mass~\cite{juanjo}.
Thus, one has the large--$N_C$ relations,
\begin{eqnarray}
\label{eq.Zpi}
&& \hspace*{-0.75cm}
Z_\pi = 1\, - \, \Frac{ 8 c_d c_m}{f^2}\, \Frac{m_\pi^2}{M_S^2}\,
+\, \Frac{64 c_d c_m^3}{f^4}\,
\Frac{m_\pi^4}{M_S^4}\, +\, \cO(m_\pi^6)\, ,
\\
&& \hspace*{-0.75cm}
2 B_0 \hat{m} = m_\pi^2 \, +\, \Frac{8 c_m (c_d-c_m)}{f^2}\,
\Frac{m_\pi^4}{M_S^2} \, +\,
\cO(m_\pi^6)\, ,
\end{eqnarray}
with $\hat{m}$ the $u$ and $d$ quark masses in the isospin limit.
The expressions for $r_i$ provided in Ref.~\cite{op61} did not take
this effect into account. Our results in
Eqs.~(\ref{r2rf})--(\ref{r6}) are fully general and allow  a simple
implementation of this correction. Thus, one gets the corrected
widths,
\begin{eqnarray}
\Gamma_V&=&\frac{g_V^2 M_V^5 \, \rho_{\rm V}^3 }{48\pi f_\pi^4}
\left[1 +\Frac{4\sqrt2 f_\chi}{g_V}\, \Frac{2 B_0 \hat{m}}{M_V^2}\right]^2\,,
\\
\label{gammas'} \Gamma_S &=& \frac{3 c_d^2 M_S^3\, \rho_{\rm S}
}{16\pi f_\pi^4} \left[ 1 \, -  \, \Frac{2m_\pi^2}{M_S^2} +
\Frac{2c_m}{c_d}\, \Frac{2 B_0 \hat{m}}{M_S^2}\right]^2\,,
\end{eqnarray}
with $f_\pi = f \, Z_\pi^{-\frac{1}{2}}$~\cite{juanjo}. The
resonance masses remain $m_\pi$ independent. From this,
one is able to recover the real parameters that provide the LECs:
\begin{eqnarray}
&&\alpha_V\, =\, \beta_V\, = \, \Frac{8\sqrt2 f_\chi}{g_V}\, -\, 6 \,
-\, \Frac{16 c_d c_m \, M_V^2}{f^2 \, M_S^2} \,,
\\
&&\gamma_V \, = \, \frac{32f_\chi^2}{g_V^2}
\, - \,  \frac{ 48 \sqrt{2} f_\chi}{g_V}
\,\left[1 +\Frac{4 c_m (c_d+c_m)\, M_V^2}{3 f^2 \, M_S^2} \right]
\nonumber
\\
&& \, \,\,\,\, \,\,  +\, 6 \left[ 1+\Frac{16 c_d c_m M_V^2}{f^2  M_S^2} +
\Frac{ 32 c_m^2 c_d(c_d+  2 c_m)
M_V^4}{3 f^4  M_S^4} \right] ,
\nonumber \\ &&
\\
&&\alpha_S\, =\, \beta_S \, = \, \Frac{4c_m}{c_d}\, -\, 6\,
-\, \Frac{16 c_d c_m}{f^2}\,,
\\
&&\gamma_S\, = \, 10\,\left[1+\Frac
{48 c_d c_m}{5 f^2} +\Frac{32 c_d^2 c_m^2}{5 f^4}\right]
\nonumber \\ &&
\qquad -\,\frac{16c_m}{c_d} \, \left[1+\Frac{2 c_d c_m}{f^2}
-\Frac{ 8 c_d^2 c_m^2}{f^4} \right]
\nonumber \\
&& \qquad  +\, \frac{4c_m^2}{c_d^2}
\, \left[1-\Frac{8 c_d c_m}{f^2}\right] \,.
\end{eqnarray}
Substituting these values in Eqs.~(\ref{r2rf})--(\ref{r6}), one
recovers the proper values for $(r_2-2 r_f),\, r_3, \, ... r_6$.
Notice that now the original parameters in
Eq.~(\ref{av})--(\ref{gs}) has gained extra terms proportional to
$c_m$ due to the scalar tadpole originated by the operators $c_m
\langle S\chi_+\rangle$. However, $r_5$ and $r_6$ remain unchanged
and only the couplings $r_f,\, r_2,\, r_3, \, r_4$ gets modified.

In order to get the value of $r_f$ (allowing the separate extraction of $r_2$),
we need the value of the $\cO(p^4)$ LECs~\cite{Ecker89}
\begin{equation}
\label{eq.L5L8}
L_5=\frac{c_dc_m}{M_S^2}\, ,  \qquad  L_8=\frac{c_m^2}{2M_S^2}\, .
\end{equation}
The wave-function renormalization in Eq.~(\ref{eq.Zpi})
provides the value of $f_\pi$ in the resonance theory under consideration.
Comparing this to the $f_\pi$ expression in $\chi$PT from Eq.~(\ref{fpi})
and using the values of $L_5$ and $L_8$ from Eq.~(\ref{eq.L5L8}),
one can extract the corresponding $\cO(p^6)$ LEC in terms of the resonance couplings:
\begin{equation}
r_f\, = \, - \, \Frac{8c_d^2c_m^2}{M_S^4}.
\end{equation}

We proceed now to a numerical comparison of  our new calculation and
the original results in Ref.~\cite{op61}, where one had
\begin{equation}\label{oldres}
r_2\, =\, 1.3\cdot 10^{-4} \, , \quad r_3\,=\, -1.7 \cdot 10^{-4}\,
, \quad r_4\, =\, -1.0\cdot 10^{-4}\, .
\end{equation}
This can be compared to our determinations
\begin{equation}\label{newres}
r_2\, =\, 18 \cdot 10^{-4} \, , \quad r_3\,=\, 0.9 \cdot 10^{-4}\, ,
\quad r_4\, =\, -1.9\cdot 10^{-4}\, ,
\end{equation}
where we took the same inputs used in Ref.~\cite{op61} to extract
the values of the LECs in Eq.~(\ref{oldres}), $f=93.2$~MeV, $g_V=0.09$, $f_{\chi}=-0.03$,
$M_V=M_\rho=770$~MeV, $c_d=32$~MeV, $c_m=42$~MeV, $M_S=983$~MeV. The
kaon and eta contributions~\cite{op61} have also been  added in
Eq.~(\ref{newres}) in order to compare with Eq.~(\ref{oldres}).
The impact of this modifications on the whole amplitude is not large since
it is an $\cO(p^6)$ effect.

Observing the scattering-lengths derived from
Ref.~\cite{op61}, we get slight shifts on the values:
\begin{eqnarray}\label{eq.delta-aIJ}
\delta a_0^0&=&0.004,  \,\,\,\,\,\,\,\,\,\,\,\,\,\,\,\qquad \delta b_0^0=0.004 ,\nonumber \\
10 \cdot \delta a_0^2&=&-0.003, \qquad 10 \cdot \delta b_0^2=-0.017 , \\
10 \cdot \delta a_1^1&=&0.001, \,\,\,\,\,\qquad  10 \cdot \delta
b_1^1=-0.003 ,\nonumber
\end{eqnarray}
given in $m_\pi$ units for the mass-dimension quantities.
Although there are large variations in the $\cO(p^6)$ LECs (especially $r_2$), we verified that
the effect on the global uncertainties in the
current scattering-length determinations~\cite{aIJ} is negligible.
Nevertheless, the lack of control on the $r_k$ avoids any improvement of the errors beyond these values even if
the accuracy in the remaining inputs is considerably increased.
Hence, from our estimate in Eq.~(\ref{eq.delta-aIJ}) we consider that it is hard
to further decrease, for instance, $\Delta a^0_0$ below $0.004$ unless our knowledge
on the resonance parameters is adequately improved.

This exercise shows that the extraction of the these couplings
requires of a very subtle analysis and a closer examination of the
resonance lagrangian.
The langrangian in Eqs.~(\ref{lv})~and~(\ref{ls}) provides only a rough
approximation and there can be more resonance contributions to the
$\cO(p^6)$ LECs beside the scalar tadpole~\cite{rcht-op6}. These
variations due to unheeded contributions just point out the level of
theoretical uncertainty that comes into play from our ignorance of
the resonance lagrangian.

We presented in this note a new method to calculate $\chi$PT low-energy constants in terms of
resonance parameters in a model independent way, without relying on any particular form of the
resonance lagrangian.
This technique provides
a convenient procedure of implementing the high and low-energy
constraints and can be useful for future studies.

\section*{Acknowledgments}

This work is supported in part by National Nature Science Foundations of China
under contract number
 10575002, 10421503 (H.Z.) and China Postdoctoral Science Foundation
 under Grant No.~20060400375 (J.J. S-C.). We would also like to
 thank Joaquim Prades for helpful suggestions which initiated this work and Heinrich Leutwyler for
 valuable comments.


\begin{thebibliography}{26}

\bibitem{lnc} Z.~H.~Guo, J.~J.~Sanz~Cillero, H.~Q.~Zheng, JHEP
06(2007) 030.


\bibitem{Ecker89}G. Ecker {\it et al.}, Nucl. Phys. {\bf B321} (1989)311.


\bibitem{spin1fields}
    G. Ecker {\it et al.}, {\it Phys. Lett.} B {\bf 223} (1989) 425.


\bibitem{KSRF}
    K.~Kawarabayashi and M.~Suzuki,
    Phys. Rev. Lett. {\bf 16} (1966) 255;
    Riazuddin and Fayazuddin,
    Phys. Rev. {\bf 147} (1966) 1071; \\
    J.~L.~Basdevant and J.~Zinn-Justin,
    Phys. Rev. D {\bf 3} (1971) 1865;
    See also, S.~Rudaz,
    Phys.~Rev. D {\bf 10}(1974)3857;
    G.~Kramer and W.~F.~Palmer,
    Phys. Rev. D {\bf 36}(1987)154.



\bibitem{Polyakov}
    Analogous predictions for the LECs can be found in
    A.~A.~Bolokhov {\it et al.},
    Phys. Rev. D {\bf 48} (1993) 3090.



\bibitem{chptweinberg}
  S. Weinberg,
  Physica {\bf 96A} (1979) 327.

\bibitem{chptoneloop}
  J. Gasser and H. Leutwyler,
  Annals Phys. {\bf 158} (1984) 142.

\bibitem{gasser85}
  J.~Gasser and H.~Leutwyler,
  {\it Nucl. Phys.} {\bf B  250} (1985) 465;


\bibitem{lhc}See also Z.~X.~Sun {\it et al.}, Mod. Phys. Lett.{\bf
A22} (2007) 711; Z.~G.~Xiao, H.~Q.~Zheng, Mod. Phys. Lett. {\bf A22}
(2007) 55;

\bibitem{Xiao00}Z.~G.~Xiao and H.~Q.~Zheng, Nucl. Phys. {\bf A 695}(2001)273.

\bibitem{martin}B.~R.~Martin, D.~Morgan and G.~Shaw, {\it Pion Pion Interactions in
Particle Physics}, Academic Press, London, 1976

\bibitem{zheng}H.~Q.~Zheng $et$ $al.$, Nucl. Phys. {\bf
A}733(2004)235;\\ Z.~Y.~Zhou and H.~Q.~Zheng, Nucl. Phys. {\bf
A755}(2006)212;\\J.~Y.~He, Z.~G.~Xiao and H.~Q.~Zheng,
 Phys. Lett. {\bf B536}(2002)59; Erratum ibid.{\bf B549}(2002)362.



\bibitem{op61}
J.~Bijnens, G.~Colangelo, G.~Ecker, J.~Gasser, M.~E.~Sainio,
 Nucl.Phys.{\bf B 508}(1997) 263, Erratum-ibid.{\bf B 517}(1998) 639.

\bibitem{juanjo} J.~J.~Sanz-Cillero,  Phys. Rev. {\bf D 70} (2004) 094033.

\bibitem{Kpi-SFF}
    M.~Jamin, J.~A.~Oller and A.~Pich,
    Nucl. Phys. B {\bf 622} (2002) 279-308.

\bibitem{f2loop} J.~Bijnens, G.~Colangelo and P.~Talavera, JHEP 05(1998) 014;
\\
J.~Bijnens, G.~Colangelo and G.~Ecker, Annals. Phys 280(2000) 100.


\bibitem{aIJ}
    G. Colangelo, J. Gasser and H. Leutwyler,
    Nucl. Phys. B {\bf 603} (2001) 125.

\bibitem{rcht-op6}
    V. Cirigliano {\it et al.},
    Nucl. Phys. B {\bf 753} (2006) 139-177.



\end{thebibliography}
\end{document}